\documentclass[10pt, final, journal, twoside]{IEEEtran}

\usepackage{upgreek} 
\usepackage{caption}
\usepackage{cite}
\usepackage{mathtools}
\usepackage{epsf}
\usepackage{commath}
\usepackage{amsmath}
\usepackage{amssymb}
\usepackage{amsfonts}
\usepackage{graphicx}
\usepackage{mwe,tikz}
\usepackage[percent]{overpic}
\usepackage{psfrag}
\usepackage{array,multirow,graphicx}
\usepackage{newunicodechar}
\usepackage{subfigure}
\newcommand{\re}{r_{\mathrm{e}}}
\newcommand{\phimax}{\phi_{\mathrm{max}}}

\newcommand{\Gammamin}{\Upupsilon_{\mathrm{min}}}

\newcommand{\gammamin}{\gamma_{\mathrm{min}}}
\newcommand{\tmin}{t_{\gamma_{\mathrm{min}}}}
\newcommand{\alphamax}{\alpha_{\mathrm{max}}}

\def\E{{\textrm{E}}}

\newtheorem{example}{Example}
\newtheorem{theorem}{Theorem}

\newtheorem{lemma}{Lemma}
\newtheorem{remark}{Remark}
\newtheorem{proposition}{Proposition}
\usepackage{scalerel}
\newtheorem{corollary}{Corollary}
\usepackage{tikz}
\usetikzlibrary{svg.path}
\newcounter{MYtempeqncnt}

\definecolor{orcidlogocol}{HTML}{A6CE39}
\tikzset{
  orcidlogo/.pic={
    \fill[orcidlogocol] svg{M256,128c0,70.7-57.3,128-128,128C57.3,256,0,198.7,0,128C0,57.3,57.3,0,128,0C198.7,0,256,57.3,256,128z};
    \fill[white] svg{M86.3,186.2H70.9V79.1h15.4v48.4V186.2z}
                 svg{M108.9,79.1h41.6c39.6,0,57,28.3,57,53.6c0,27.5-21.5,53.6-56.8,53.6h-41.8V79.1z M124.3,172.4h24.5c34.9,0,42.9-26.5,42.9-39.7c0-21.5-13.7-39.7-43.7-39.7h-23.7V172.4z}
                 svg{M88.7,56.8c0,5.5-4.5,10.1-10.1,10.1c-5.6,0-10.1-4.6-10.1-10.1c0-5.6,4.5-10.1,10.1-10.1C84.2,46.7,88.7,51.3,88.7,56.8z};
  }
}

\newcommand\orcidicon[1]{\href{https://orcid.org/#1}{\mbox{\scalerel*{
\begin{tikzpicture}[yscale=-1,transform shape]
\pic{orcidlogo};
\end{tikzpicture}
}{|}}}}

\usepackage{hyperref} 
\usepackage{breakurl}

\begin{document}

\title{{\LARGE Statistical Modeling for Accurate Characterization of Doppler Effect in LEO-Terrestrial Networks }}
\author{Islam~M.~Tanash$^{ \orcidicon{0000-0002-9824-6951}}$, Risto Wichman$^{\orcidicon{0000-0002-5261-5037}},$~\IEEEmembership{Senior Member,~IEEE}, and Nuria González-Prelcic$^{\orcidicon{0000-0002-0828-8454}},$~\IEEEmembership{Fellow,~IEEE}%
\thanks{Islam M. Tanash is with the Department of Electrical Engineering, Prince Mohammad Bin Fahd University, Al Khobar 31952, Saudi Arabia (e-mail: itanash@pmu.edu.sa).}

\thanks{Risto Wichman is with the Department of Information and Communications Engineering, Aalto University, 00076 Espoo, Finland (e-mail: risto.wichman@aalto.ﬁ).}

\thanks{Nuria Gonz´alez-Prelcic is with the Department of Electrical and Computer
Engineering, University of California San Diego, La Jolla, CA 92093 USA
(e-mail: ngprelcic@ucsd.edu).}%
\thanks{Digital Object Identifier X}}
\maketitle

\begin{abstract}
Low Earth Orbit (LEO) satellite communication is a promising solution for global wireless coverage, especially in underserved and remote areas. However, the high relative velocity of LEO satellites induces significant Doppler shifts that disrupt subcarrier orthogonality and degrade multicarrier system performance. While the common time-varying Doppler shift can be compensated relative to a reference point, the residual differential Doppler across users within the coverage cell remains a significant challenge, causing severe intercarrier interference.
This paper presents a generalized analytical framework for characterizing both the Doppler shift magnitude and the differential Doppler in LEO systems. Unlike prior works limited by flat-Earth assumptions or specific orbital configurations, our model incorporates Earth’s curvature and supports arbitrary elevation angles. Using spherical geometry, we derive closed-form expressions for Doppler shift based on the central angle between the satellite and ground users. We further provide a statistical characterization of both the Doppler shift magnitude and the differential Doppler in terms of their cumulative distribution function (CDF) and probability density function (PDF) for uniformly distributed users within a spherical cap cell. Additionally, we derive a tight upper bound for the Doppler shift CDF and an exact expression for the maximum differential Doppler experienced across the coverage region.
To mitigate intra-cell Doppler variation, we implement a user clustering technique that partitions the coverage area based on a Doppler disparity threshold into spherical sub-cells, ensuring compliance with 3GPP tolerances. Extensive simulations over realistic satellite constellations validate our analysis and reveal the impact of altitude, beamwidth, and satellite-user geometry on Doppler behavior.
\end{abstract}
\begin{IEEEkeywords}
Low Earth orbit (LEO) satellites, Doppler shift, stochastic
geometry.
\end{IEEEkeywords}

\section{Introduction}
\label{sec:Introduction}
There has been an increasing interest in using satellite communications to achieve global wireless coverage that complements and extends terrestrial networks. Recent studies have focused on low Earth orbit (LEO) satellite communication systems due to their fewer restrictions, particularly in terms of power consumption and latency, compared to their medium Earth (MEO) and Geostationary orbit (GEO) counterparts~\cite{leo-general}. LEO satellite communication is an excellent solution for providing connectivity in underserved areas lacking infrastructure, such as rural and remote areas. Its importance is further emphasized by 3GPP's integration of non-terrestrial networks into the 5G framework~\cite{TR38.821}. LEO satellites have the potential to solve the problem of significant delays in communication links that are common in GEO satellites. However, they face various challenges, with high mobility being a prominent issue that can cause an increased Doppler shift, which in turn can hinder their performance~\cite{servey_doppler}.

In the context of satellite communication, the Doppler shift is the change in the frequency of the electromagnetic waves due to the relative velocity between the satellite and the ground users. Frequency compensation techniques may partially eliminate the Doppler shift~\cite{compen4,compensate,compensation2,compens3}. However, the residual Doppler shift, also known as the differential Doppler shift, persists among users within a coverage area due to their varied geographical locations, which lead to variations in their velocity vectors relative to the transmitting satellite. Differential Doppler shift can cause intercarrier interference and potentially affect the subcarrier orthogonality and the performance of a multicarrier system. Therefore, predicting the Doppler shift in the satellite's coverage area is necessary to provide reliable service in satellite communications~\cite{daher_ref_1,daher_ref_2,ali, compensation1}.

In~\cite{daher_ref_1}, the authors examined the impact of Doppler shift on LEO satellite communication systems, focusing specifically on its implications for carrier recovery at the receiver. However, the analysis is restricted to a simplified scenario involving circular LEO satellites confined to the equatorial plane, with ground receivers positioned exclusively along the equator's ground trace. In \cite{daher_ref_2}, the authors derived analytical expressions for the Doppler shift and the elevation angle for satellites moving on elliptical orbits as a function of the time at which the satellite is closest to the ground station, i.e., the perigee time. A more general study was conducted in \cite{ali}, in which the authors employed spherical geometry to derive accurate expressions for Doppler shift and the satellite's visibility window duration as a function of the maximum elevation angle between the ground terminal and the satellite.
In~\cite{compensation1}, information on the instantaneous satellite position and the estimated terminal position was used to predict the continuous Doppler shift in closed form in terms of geographical parameters, whereas in \cite{simulator}, the same information was used to design a simulator that can calculate the Doppler shift for any satellite orbit. However,~\cite{compensation1} and \cite{simulator} rely on precise real-time location data for Doppler compensation, making them vulnerable to location errors and imposing high computational demands on user equipment.

Building upon~\cite{ali}, several works tackled the problem of characterizing the differential Doppler shift, which exists among users distributed within the coverage area of a specific LEO satellite~\cite{uplink,dop_iot,main}. In particular, the authors in~\cite{uplink,dop_iot} presented an upper bound for the differential Doppler in a Narrowband Internet of Things (NB-IoT) cell served by a LEO satellite. To reduce the maximum Doppler shift, they proposed dividing the coverage area into smaller regions, ensuring that the differential Doppler within each region is below a predefined threshold, and proposed time-domain and frequency-domain group scheduling strategies to enable reliable carrier recovery at the base station.

A general characterization of the Doppler shift was presented in~\cite{main} exploiting stochastic geometry tools that substitute the deterministic network topology of ground users. Ground users within a coverage area were clustered using Matern cluster process (MCP) and the distribution of the Doppler shift magnitude experienced by a randomly selected user within a cluster was identified. However, the proposed analytical framework is based on a ﬂat-Earth approximation, thereby disregarding the curvature of the Earth. This oversimplified approach leads to less accurate analyses, particularly for larger clusters. Moreover, the presented distribution of the Doppler shift magnitude within a cluster lacks generality. It is specifically tailored for a particular scenario when the cluster containing the users lies on the ground track, the path formed by successive subsatellite points directly beneath the satellite’s orbit, with the maximum elevation angle set to $\frac{\pi}{2}$. This case is subsequently validated as an upper bound for situations where clusters containing users are located outside the orbital plane. Nevertheless, this upper bound is very loose and does not lead to reliable analysis.

Motivated by the above limitations, in this paper we aim to accurately characterize the magnitude of the Doppler shift, overcoming the limitations highlighted in~\cite{main} and accommodating any value of the maximum elevation angle. A significant improvement in accuracy is further achieved by incorporating the curvature of the Earth's surface.
The contributions of this paper can be summarized as follows:
\begin{itemize}
\item We present a comprehensive analytical framework for modeling Doppler shifts in LEO satellite systems using spherical geometry that accurately accounts for the Earth's curvature. Within this framework, we derive closed-form expressions for the cumulative distribution function (CDF) and probability density function (PDF) of the Doppler shift magnitude experienced by a randomly located user within a satellite beam. This enables a more precise representation of Doppler effects compared to traditional flat-Earth approximations, especially under wide beam coverage and low elevation angles.

\item We extend the framework to characterize the differential Doppler shift across users within a satellite footprint by deriving its CDF and PDF. This provides a detailed statistical understanding of Doppler variations across the satellite's coverage area.

\item We derive a tight upper bound for the CDF of the Doppler shift magnitude, offering significantly improved accuracy over the bound in~\cite{main}.

\item We derive an exact expression for the maximum differential Doppler experienced within a satellite footprint, assuming uniformly distributed users under a common visibility window. 

\item We implement a Doppler-aware user clustering technique that partitions the coverage region into smaller spherical caps based on a tunable Doppler disparity threshold. This reduces intra-cluster Doppler spread and ensures that differential Doppler remains within the acceptable limits defined by 3GPP standards (e.g., 950 Hz).

\item We validate the proposed framework through extensive system-level simulations using realistic LEO constellation models (e.g., Walker-star). Results confirm the high accuracy of our analytical expressions and demonstrate the impact of altitude, cell's size, and satellite position on Doppler dynamics.
\end{itemize}


\section{System Model}
\label{sec:system model}
\begin{figure}
    \centering
      \includegraphics[trim=7.0cm 2.5cm 8cm 0.9cm, clip=true, width=.49\textwidth ]{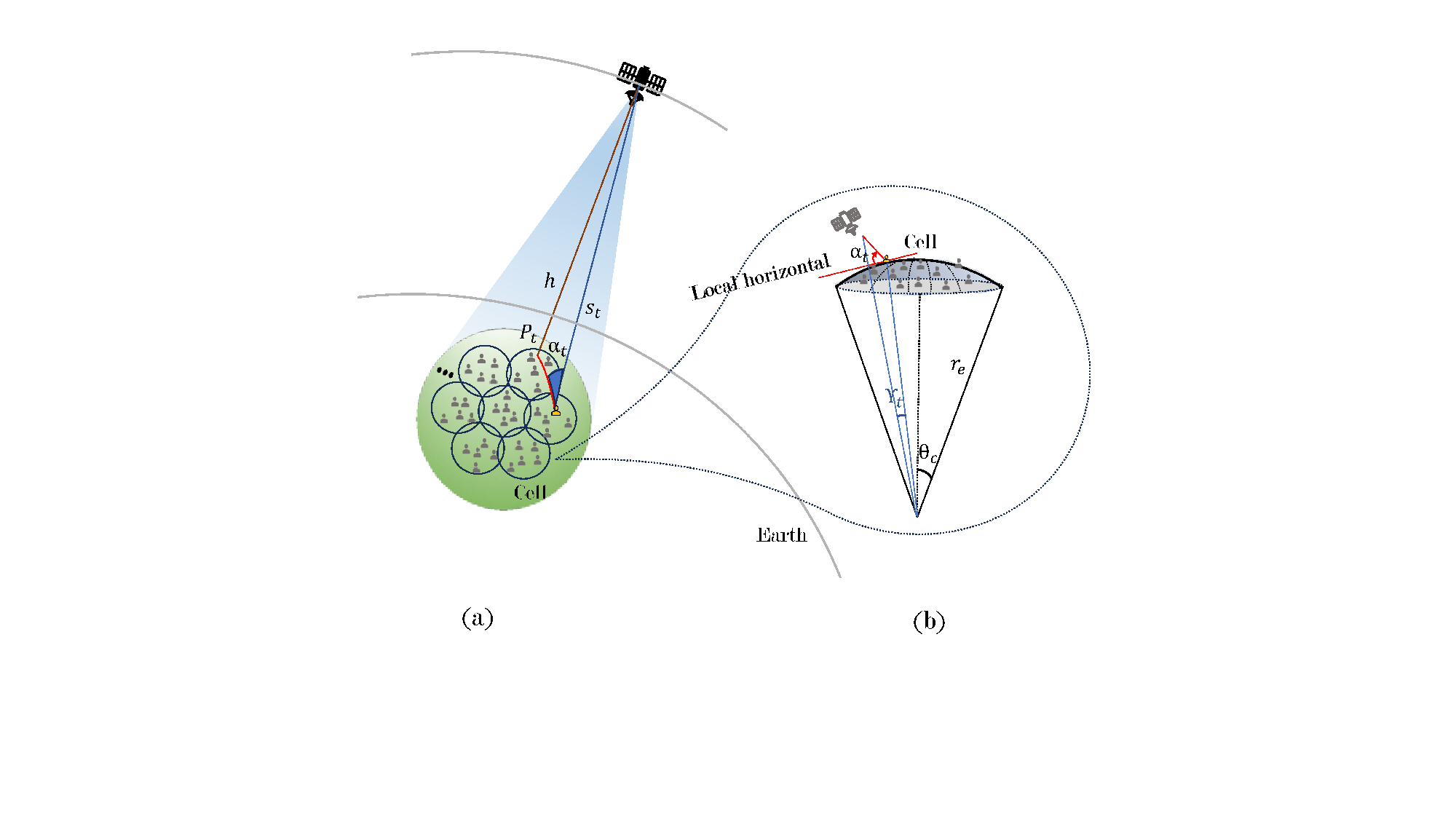}
    \caption{Illustration of a satellite communication system with multiple spot beams. (a) A satellite at altitude $h$ creates multiple spot beams, each covering a distinct cell on the Earth's surface. (b) Geometric representation of a spot beam forming a cell defined by a central angle $\theta_c$ from the cell's center to its edge. The satellite's projection onto the Earth's surface at time $t$ is denoted by $P_t$. The slant distance to the UE is $s_t$, with elevation angle $\alpha_t$, and corresponding central angle $\Upupsilon_t$.}
    \label{fig:system_model}
\end{figure}

The system under study includes a constellation of LEO satellites, each with multiple Earth-ﬁxed spot beams, as shown in Fig.~\ref{fig:system_model}(a). Generally, LEO satellites cover the target area with multiple independent spot beams, with frequency resources divided into multiple frequency bands, allowing multiple beams to operate at the same frequency to achieve higher system capacity and flexibility compared to single-beam antenna systems~\cite{multi_beam_letter}. Circular spot beams, when directed onto the surface of Earth, generate spherical cap cells that are encompassed by an angle $2\theta_c$, where $\theta_c$ denotes the central angle from the cell's center $C$ to its edge, as depicted in Fig.~\ref{fig:system_model}(b). These cells represent the footprint of the spot beams on the Earth's surface, taking into account the curvature of Earth, and are defined as $S(C,\theta_c)$, where $C$ is the caps' center.
We assume user equipments (UEs) are distributed uniformly at random within the spherical cap cell, each with a spherical area of $A_c=2\,\pi\,\re^2\,(1-\cos(\theta_c))$.

The projection of the serving satellite at an altitude $h$ onto the Earth's surface at time $t$ is denoted by $P_t$. The slant distance between the satellite and UE is denoted by $s_t$, with the elevation angle to the satellite as $\alpha_t$ and the corresponding central angle as $\Upupsilon_t$. The maximum possible elevation angle between the UE and the satellite in the circular orbit is denoted by $\alphamax$, and the corresponding central angle by $\Gammamin$. The minimum central angle from the cell's center to the nearest point on the serving orbit is denoted by $\mu_{_{\Gammamin}}$, while $\theta_v$ denotes the central angle between $P_t$ and the cell's center $C$.

The spatially distributed UEs will experience different levels of Doppler shift due to their different locations and elevation angles with respect to the traversing serving satellite. 
Ground Gateways exchange critical control commands and operational data with satellites, including satellite position and velocity, via feeder links which are backbone elements in satellite communication~\cite{feeder_link}. Therefore, the Gateway can calculate the common time-varying Doppler shift relative to a reference point within the cell, which can then be pre-compensated in the downlink or post-compensated in the uplink at the Gateway~\cite{dop_iot}. This leads to residual Doppler shifts at the UEs, which grow with distance from the reference point where the Doppler is nullified. As a result, differential Doppler becomes the main source of performance degradation, requiring precise characterization for effective analysis and system optimization.

In this paper, we make the following assumptions:

\textit{Assumption $1$ (Reference Instant):}
The instant when the UE has its minimum central angle to the satellite is set as the reference time instant where the UE has zero Doppler shift.

\textit{Assumption $2$ (Satellite Trajectory):}
Due to the relatively small visibility window for a LEO satellite compared to its orbital period, the satellite's trajectory can be approximated by a great circle arc in the Earth-centered fixed (ECF) frame. 

\textit{Assumption $3$ (Angular Velocity):}
The variation in the angular velocity of the satellite in the ECF frame with latitude due to the rotation of the Earth is very small for most LEO circular orbits, and can, therefore, be approximated by a constant using~\cite[Eq.~9]{ali} $\omega_F=\omega_s-\omega_E\,\cos(i)$, where $\omega_s$ and $\omega_E$ indicate respectively the satellite's and Earth's angular velocities in the Earth-centered inertial (ECI) coordinate system. These velocities are typically constant in the ECI frame with $\omega_s=\sqrt{\frac{\mu}{(\re+h)^3}}$, where $\mu$ is the standard gravitational parameter of Earth, $\re$ is the Earth's radius, and $i$ is the constellation's inclination.  


\section{Doppler Characterization}
\begin{figure*}[h]
    \centering
      \includegraphics[trim=2cm .5cm 2cm .5cm, clip=true, width=.9\textwidth ]{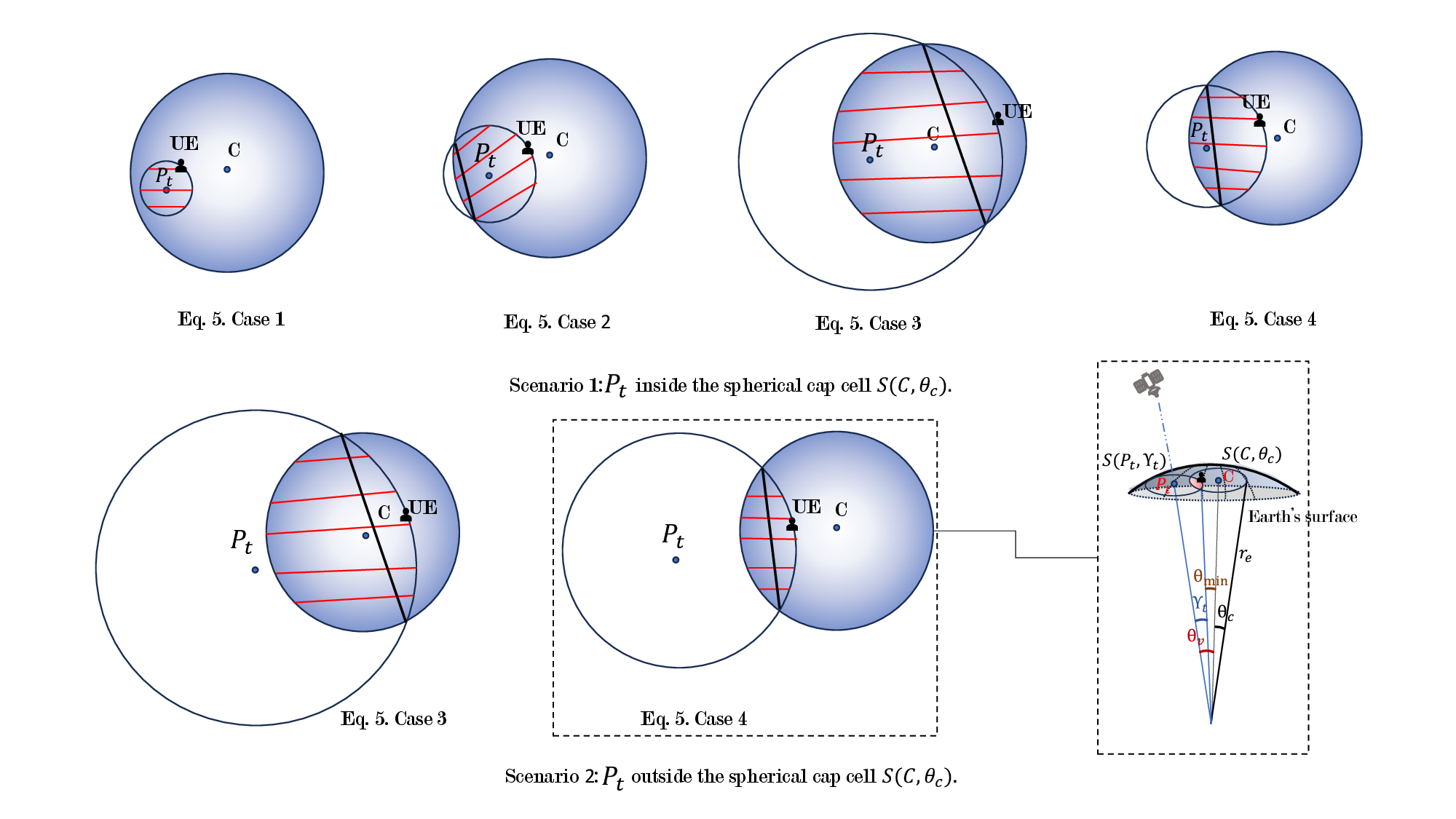}
 \caption{Illustration of the possible cases for the location of the projection point $P_t$ relative to the spherical cap coverage cell $S(C,\theta_c)$ and the UE location, as described in (\ref{eq:central_cdf}). Scenario 1 and Scenario 2 represent $P_t$ inside and outside $S(C,\theta_c)$, respectively. The rightmost subfigure shows the geometric relation between the satellite, the spherical cap coverage cell $S(C,\theta_c)$ and $S(P_t,\Upupsilon_t)$, together with the involved angles.}

    \label{fig:cases}
\end{figure*}

\begin{figure*}[t]
\setcounter{MYtempeqncnt}{\value{equation}}
\setcounter{equation}{4}
\begin{equation}
\resizebox{1.0\hsize}{!}{$\begin{aligned}\label{eq:central_cdf}
F_{\Upupsilon_t}(\gamma_t,\theta_c\mid\theta_v)&=\begin{cases}
\frac{1-\cos(\gamma_t)}{1-\cos(\theta_c)},&\text{Case 1: } 0 \leq \gamma_t \leq {\theta_c-\theta_v}, \\ 
\frac{1-\cos(\gamma_t)}{1-\cos(\theta_c)}+\frac{1}{2(1-\cos(\theta_c))}\,\left[V_{\theta_v+\theta_{\min },\theta_c}(\gamma_t) -V_{\theta_{\min },\gamma_t}(\gamma_t)\right], & \text{Case 2: } \theta_c > \theta_v\,\, \& \,\, \cos \left(\gamma_t\right) \cos \left(\theta_v\right) \geq \cos \left(\theta_c\right),  \\
1+\frac{1}{2(1-\cos(\theta_c))}\,\left[V_{\theta_v+\theta_{\min },\gamma_t}(\gamma_t) -V_{\theta_{\min },\theta_c}(\gamma_t)\right], & \text{Case 3: } \theta_v < \gamma_t \leq \theta_c+\theta_v \,\, \&\,\,\cos \left(\gamma_t\right) \leq \cos \left(\theta_c\right) \cos \left(\theta_v\right),\\
\frac{1}{2(1-\cos(\theta_c))}\,\left[ V_{\theta_{\min },\theta_c}(\gamma_t)+V_{\theta_v-\theta_{\min },\gamma_t}(\gamma_t)\right], & \text{Case 4: otherwise.} 
\end{cases}
\end{aligned}$}
\end{equation}
\setcounter{equation}{\value{MYtempeqncnt}}
\hrulefill
\end{figure*}

\label{sec:doppler characterization}
In this section, we leverage the analytical findings obtained in~\cite{ali} for the Doppler shift at an observation point on Earth. By focusing on central angles, we establish the primary characterization of the Doppler shift, which is then employed to derive highly accurate distributions for the Doppler magnitude, differential Doppler, and maximum Doppler within a cell, considering the curvature of the Earth.

\subsection{Doppler Shift Equation}
In~\cite[Eq.~5]{ali}, the Doppler equation is formulated as a function of the angular distance at a specific time and maximum elevation angle while in~\cite[Eq.~2]{main}, it is formulated as a function of time, elevation angle at a specific time, and maximum elevation angle. Using these formulations as a basis, we derive the Doppler shift as a function of the central angle and maximum central angle in the following Lemma, which form the base for the subsequent analysis.   

\begin{lemma}
\label{lemma:doppler_shift}
\textit{Let $\Upupsilon_t\in[0,\phimax]$ represent the central angle from a stationary UE on the Earth’s surface to a circular orbit LEO satellite at time $t$, where $\phimax=\arccos\big(\frac{\re}{{ \re+h}}\big)$, derived from the law of cosines. The Doppler shift magnitude, denoted by $\delta_t$, resulting from satellite mobility can be analytically expressed relative to time $t=\tmin$, when the UE observes the minimum central angle $\gammamin$, as
\begin{align}
\label{eq:doppler_s}
\delta_t &= \rho \, \frac{\sqrt{\cos^2(\Gammamin) - \cos^2(\Upsilon_t)}}{\sqrt{1 + k^2 - 2k\cos(\Upupsilon_t)}},
\end{align} 
where $\rho=\frac{f_o\,\re \,\omega_f}{c}$ with $f_o$ denoting the carrier frequency, $c$ the speed of light and $k = \frac{\re}{\re + h}$.}
\end{lemma}

\begin{IEEEproof}
From~\cite[Eq.~2.35]{book_elav}, we substitute $\cos(\alpha_t)={\sin \left(\Upupsilon_t\right)}\big{/}\sqrt{1+k^2-2k \cos (\Upupsilon_t)}$ into the absolute form of the Doppler shift in~\cite[Eq.~2]{main}. Additionally, we utilize $\omega_F\left(t-\tmin\right)=\cos^{-1}\left(\frac{\cos (\Upupsilon_t)}{\cos(\Gammamin)}\right)$ from algebraic manipulation of~\cite[Eq.~10]{ali}, and apply the trigonometric identity $\sin(\cos^{-1}(x))=\sqrt{1-x^2}$. This result is then substituted into~\cite[Eq.~3]{main}. We further use the identity $\sin(x)^2+\cos(x)^2=1$ and utilize $\Gammamin=\cos ^{-1}\big(k \cos \left(\alphamax\right)\big)-\alphamax$ from~\cite[Eq.~4]{ali} and substitute it in~\cite[Eq.~4]{main}.
\end{IEEEproof}

\begin{remark}
Equations (\ref{eq:doppler_s}) pertains to the Doppler shift experienced by individual users within the cell. Consequently, we note that various UEs not only have different central angles to satellites but also distinct minimum central angles. Hence, both $\Upupsilon_t$ and $\Gammamin$ emerge as random variables, which require the derivation of their statistical characterization to accurately define the differential Doppler shift within the cell.
\end{remark}

\begin{remark}
The expression in~(\ref{eq:doppler_s}) indicates the magnitude of the Doppler shift, while its sign can be determined by the direction of movement of the LEO satellite relative to the UE. In the context of this paper, when the satellite approaches the point on the serving orbit where the UE observes the minimum central angle (maximum elevation angle), the Doppler shift is positive. Conversely, if the satellite moves away from this point, the sign is negative. It is important to note that when the satellite aligns with the point on the serving orbit where the UE observes the minimum central angle, the Doppler shift at the UE becomes zero.
\end{remark}

\begin{corollary}
\label{corol:special_case_1}
\textit{When the UE lies on the satellite’s ground track, which represents the locus of consecutive subsatellite points where $\Gammamin=0$, the Doppler shift magnitude in (\ref{eq:doppler_s}) simplifies to 
\begin{align}
\label{eq:doppler_s_simplified}
{\delta}_t (\Upupsilon_t) &=\rho\, \frac{\sin(\Upupsilon_t)}{\sqrt{1+k^2-2k\cos(\Upupsilon_t)}}.
\end{align} }
\end{corollary}

\begin{lemma}
\textit{The Doppler shift magnitude relative to time $t=\tmin$, when the UE observes the minimum central angle $\Gammamin$, can be approximated by
\begin{align}
\label{eq:doppler_s_approx}
\widetilde{\delta_t}(\Upupsilon_t) &=\rho\, \,\sqrt{\frac{\Upupsilon_t^2-\Gammamin^2}{\big(\frac{h}{\re+h}\big)^2+\big(\frac{\re}{\re+h}\big)\Upupsilon_t^2}}.
\end{align} }
\end{lemma}
\begin{IEEEproof}
The Doppler shift in~(\ref{eq:doppler_s}) can be approximated by leveraging the small-angle approximation principle, given that $\{\Upupsilon_t, \Gammamin\} \leq \phimax$, where $\phimax$ itself is relatively small, e.g., with $h=1000$ km, $\phimax=0.527$.
Using the trigonometric identity $\sin^2(x) = 1 - \cos^2(x)$ and the small-angle approximations $\cos(x) \approx 1 - \frac{x^2}{2}$ and $\sin(x) \approx x$, followed by algebraic simplifications, we get (\ref{eq:doppler_s_approx}).
\end{IEEEproof}

\begin{corollary}
\label{corol:special_case_1}
\textit{When the UE lies on the satellite’s ground track, at which $\Gammamin=0$, the Doppler shift magnitude approximation in (\ref{eq:doppler_s_approx}) simplifies to
\begin{align}
\label{eq:doppler_s_approx_theta_min_0}
\widetilde{\delta_t}(\Upupsilon_t) &=\rho\,(\re+h)\, \,{\frac{\Upupsilon_t}{\sqrt{h^2+\re{(\re+h)}\Upupsilon_t^2}}}.
\end{align} }
\end{corollary}
\subsection{Central Angle Distribution}
\label{subs:angle_distributions}
In the following lemmas, we will introduce the CDF and PDF of the necessary central angles needed to eventually derive the Doppler magnitude distribution within a cell given the satellite location relative to a reference point at a certain time instant.

\begin{lemma}
\label{lemma:central_cdf}
\textit{The CDF of the central angle,$\Upupsilon_t$, between a serving satellite and the random UEs that are distributed uniformly in a spherical cap cell, is derived as in (\ref{eq:central_cdf}) at the top of this page, \stepcounter{equation} 
for which }
\begin{align}
\theta_{\min }(\gamma_t)=\begin{cases}\arctan \left(\frac{1}{\tan \left(\theta_v\right)}-\frac{\cos \left(\theta_c\right)}{\cos \left(\gamma_t\right) \sin \left(\theta_v\right)}\right),
 \text{Case 2, }\\
\arctan \left(\frac{1}{\tan \left(\theta_v\right)}-\frac{\cos \left(\gamma_t\right)}{\cos \left(\theta_c\right) \sin \left(\theta_v\right)}\right), \text{Case 3, }\\
\arctan \left(\frac{\cos \left(\gamma_t\right)}{\cos \left(\theta_c\right) \sin \left(\theta_v\right)}-\frac{1}{\tan \left(\theta_v\right)}\right), \text{Case 4, }\\
\end{cases}
\end{align}
\begin{align}
\label{eq:v_original}
 V_{u, v}(\gamma_t)=:&\int_{u(\gamma_t)}^{v(\gamma_t)} \sin (\phi)\, I_{1-\left(\frac{\tan \left(u(\gamma_t)\right)}{\tan (\phi)}\right)^2}\left(\frac{1}{2}, \frac{1}{2}\right) \mathrm{d} \phi
\end{align}
\textit{with $I_x(\cdot,\cdot)$ being the regularized incomplete beta function.}
\end{lemma}
\begin{IEEEproof}
Consider a spherical cap cell of width $2\theta_c$. Let $P_t$ denote the satellite projection, which falls either within the cell, represented as Scenario 1, or outside it, represented as Scenario 2, while a UE is positioned inside the cell.
The CDF of the central angle between the satellite (or its projection) and the UE, $F_{\Upupsilon_t}(\gamma_t) = \operatorname{Pr}(\Upupsilon_t \leq \gamma_t)$, is defined as the area of intersection between two spherical caps, namely $S(P_t,\Upupsilon_t)$ and $S(C,\theta_c)$, divided by $2\pi \re^2 (1-\cos(\theta_c))$, where $P_t$ and $C$ are the caps' centers, and $\Upupsilon_t$ and $\theta_c$ are their central angles. When $P_t$ is inside $S(C,\theta_c)$, four cases can arise as illustrated by scenario 1 in Fig.~\ref{fig:cases}, and the intersection area can be calculated as follows:\\
Case 1: $S(P_t,\gamma_t)$ is entirely located inside $S(C,\theta_c)$, the intersection area is the area of the spherical cap $S(P_t,\gamma_t)$, given by $2\pi\re^2(1-\cos(\gamma_t))$.\\
Case 2: The boundary line is on the left side of both centers $C$ and $P_t$, the intersection area is provided in~\cite[Case 6]{curved_cells}.\\
Case 3: The boundary line is on the right side of both centers $C$ and $P_t$, the intersection area is provided in~\cite[Case 7]{curved_cells}.\\
Case 4: The boundary line is between the centers $C$ and $P_t$, the intersection area is provided in~\cite[Case 8]{curved_cells}.\\
On the other hand, when $P_t$ is outside $S(C,\theta_c)$, cases 3 and 4 can occur, generating the similar intersection areas as the equivalent ones in scenario 1, as depicted in Fig.~\ref{fig:cases}.  
For a given satellite, its projection $P_t$ at a specific time instant mostly lie outside the cell in which the UE is located, i.e., outside $S(C,\theta_c)$. As a result, Scenario 2 becomes the most relevant case in this study.




\end{IEEEproof}

\begin{figure*}[t]
\setcounter{MYtempeqncnt}{\value{equation}}
\setcounter{equation}{8}
\begin{equation}
\resizebox{.95\hsize}{!}{$\begin{aligned}\label{eq:central_angle_pdf}
f_{\Upupsilon_t}(\gamma_t,\theta_c\mid\theta_v)&=\begin{cases}
\frac{\sin(\gamma_t)}{1-\cos(\theta_c)},&\text{Case 1: } 0 \leq \gamma_t \leq \theta_c-\theta_v, \\ 
\frac{\sin(\gamma_t)}{1-\cos(\theta_c)}+\frac{1}{2(1-\cos(\theta_c))}\,\big[ {\widetilde{V}'}_{\theta_v+\theta_{\min },\theta_c}(\gamma_t)-{\widetilde{V}'}_{\theta_{\min },\gamma_t}(\gamma_t)\big], & \text{Case 2: } \theta_c > \theta_v\,\, \& \,\, \cos \left(\gamma_t\right) \cos \left(\theta_v\right) \geq \cos \left(\theta_c\right),\\ 
\frac{1}{2(1-\cos(\theta_c))}\,\big[{\widetilde{V}'}_{\theta_v+\theta_{\min },\gamma_t}(\gamma_t) -{\widetilde{V}'}_{\theta_{\min },\theta_c}(\gamma_t)\big], & \text{Case 3: } \theta_v < \gamma_t \leq \theta_c+\theta_v \,\, \&\,\,\cos \left(\gamma_t\right) \leq \cos \left(\theta_c\right) \cos \left(\theta_v\right),\\
\frac{1}{2(1-\cos(\theta_c))}\,\big[ {\widetilde{V}'}_{\theta_{\min },\theta_c}(\gamma_t)+{\widetilde{V}'}_{\theta_v-\theta_{\min },\gamma_t}(\gamma_t)\big], & \text{Case 4: otherwise.} 
\end{cases}
\end{aligned}$}
\end{equation}
\setcounter{equation}{\value{MYtempeqncnt}}
\hrulefill
\end{figure*}

\begin{figure*}[t]
\setcounter{MYtempeqncnt}{\value{equation}}
\setcounter{equation}{21}
\begin{equation}
\resizebox{.95\hsize}{!}{$\begin{aligned}\label{eq:PDF_DOPPLER}
& f_{\delta}(s\mid\Gammamin= \mu_{_{\Gammamin}}){=} -\frac{f_{\Upupsilon_t}\Bigg(\cos^{-1}{\bigg(\frac{s^2 k + \sqrt{s^4 k^2 - \rho^2 \big(s^2 (1 + k^2) - \rho^2 \cos^2(\mu_{_{\Gammamin}})\big)}}{\rho^2}\bigg)},\theta_c\mid\theta_v\Bigg)}{\sqrt{1 - \left( \frac{s^2 k + \sqrt{s^4 k^2 - \rho^2 s^2 (1 + k^2) + \rho^4 \cos^2(\mu_{_{\Gammamin})})}}{\rho^2} \right)^2 }} 
\cdot \frac{2s k + \frac{s (2 s k^2 - \rho^2 (1 + k^2))}{\sqrt{s^4 k^2 - \rho^2 s^2 (1 + k^2) + \rho^4 \cos^2(\mu_{_{\Gammamin})}}}}{\rho^2}
\end{aligned}$}
\end{equation}
\setcounter{equation}{\value{MYtempeqncnt}}
\hrulefill
\end{figure*}
\begin{proposition}
\label{prob:V_approx}
\textit{The expression $V_{u, v}(\gamma_t)$ in (\ref{eq:v_original}), which lacks a closed-form representation, can be effectively approximated using elementary functions as 
\begin{equation}
\resizebox{1.0\hsize}{!}{$\begin{aligned}\label{eq:v_approx}
&{\widetilde{V}}_{u, v}(\gamma_t)=\sum_{n=0}^{\infty}\frac{{\left(\frac{1}{2}\right)_n}}{{ n! \, B(\frac{1}{2}, \frac{1}{2})}}\Bigg[-\frac{{u^4(\gamma_t) \big(n-\frac{1}{2}\big) }}{ {4v^2(\gamma_t)}}
 \, _3F_2\Big(1,1,\frac{3}{2}-n;2,3;\frac{u^2(\gamma_t)}{v^2(\gamma_t)}\Big)\\
& + \frac{1}{2} \Bigg(\frac{v^2(\gamma_t)}{\big(n+\frac{1}{2}\big) }+ u^2(\gamma_t) \bigg(-1  + E+ 2  \log(u(\gamma_t))- 2\log(v(\gamma_t))\\
&  + \psi^{(0)}\Big(n+\frac{1}{2}\Big)\bigg)\Bigg)\Bigg],
\end{aligned}$}
\end{equation}
where $(\frac{1}{2})_n$ is the  Pochhammer symbol, $E \approx 0.577216$ is the Euler--Mascheroni constant, ${}_3F_2(\cdot,\cdot,\cdot;\cdot,\cdot;\cdot)$ is the generalized hypergeometric function, and $\psi^{(0)}(\cdot)$ is the $0$th polygamma function~\cite{numericalbook}. It is important to mention that truncating the infinite series to only a few terms can achieve high accuracy, e.g., with $n=2$, the approximation magnitude order of accuracy is $(10^{-6})$. }
\end{proposition}
\begin{IEEEproof}
Using the series expansion of $I_x(\cdot,\cdot)$ stated in \cite{incomplete_beta}, and the approximation $\sin (\phi)\approx \phi$, we evaluate the resulting integral to derive (\ref{eq:v_approx}). 
\end{IEEEproof}

\begin{figure}[h!]
    \centering
     \includegraphics[trim=.1cm .1cm .1cm .1cm, clip=true, width=0.48\textwidth ]{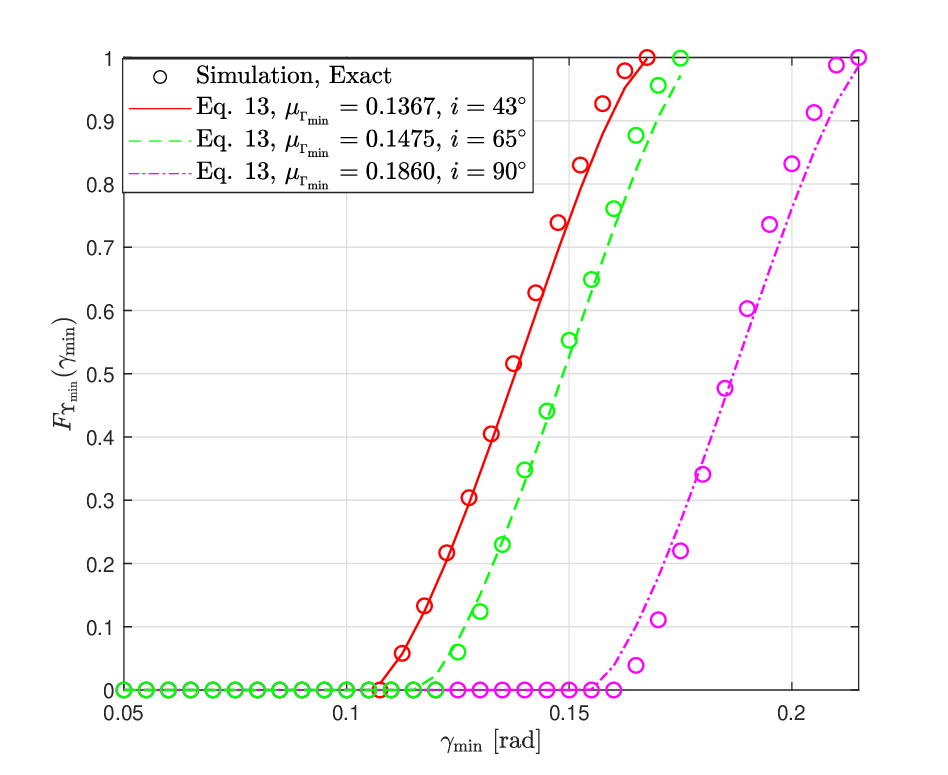}
\caption{Validation of Lemma~\ref{lemma:central_cdf}: Comparison between the exact CDF of $\Gammamin$ obtained via Monte-Carlo simulations and the proposed approximation in (\ref{eq:mean_cdf}).}
    \label{fig:mingamma}
\end{figure}

\begin{corollary}
\label{corol:central_angle_pdf}
\textit{The PDF of the central angle, $\Upupsilon_t$, between a
serving satellite and the random UEs that are distributed
uniformly in a spherical cap cell, is derived in (\ref{eq:central_angle_pdf}) at the beginning of the following page\stepcounter{equation}, for which ${\widetilde{V}'}_{u, v}(\gamma_t)$ is the derivative of ${\widetilde{V}}_{u, v}(\gamma_t)$ given in (\ref{eq:v_approx}) and can be calculated as \begin{equation}
\resizebox{1.0\hsize}{!}{$\begin{aligned}\label{eq:v_approx_der}
{\widetilde{V}'}_{u, v}(\gamma_t)=&\sum_{n=0}^{\infty}\frac{\left(\frac{1}{2}\right)_n}{\pi n!} \Bigg[-\frac{\left(n-\frac{1}{2}\right) u^4(\gamma_t) \left(\frac{2 u(\gamma_t) u'(\gamma_t)}{v^2(\gamma_t)}-\frac{2 u^2(\gamma_t) v'(\gamma_t)}{v^3(\gamma_t)}\right) }{4 v^2(\gamma_t)}\\
&\times\, _3F_2^{(0,0,0,0,0,1)}\bigg(1,1,\frac{3}{2}-n;2,3;\frac{u^2(\gamma_t)}{v^2(\gamma_t)}\bigg)+\Big(n-\frac{1}{2}\Big)\\
&\times \,_3F_2\Big(1,1,\frac{3}{2}-n;2,3;\frac{u^2(\gamma_t)}{v^2(\gamma_t)}\bigg)\bigg(-\frac{u^3(\gamma_t) u'(\gamma_t)}{v^2(\gamma_t)}+\frac{u^4(\gamma_t) v'(\gamma_t)}{2\,v^3(\gamma_t)} \bigg)\\
&+  u(\gamma_t) u'(\gamma_t) \bigg(2 \log (u(\gamma_t))-2 \log (v(\gamma_t))+\psi ^{(0)}\Big(n+\frac{1}{2}\Big)+E-1\bigg)\\
&+\frac{v(\gamma_t) v'(\gamma_t)}{(n+\frac{1}{2})}+\frac{1}{2}u^2(\gamma_t) \left(\frac{2 u'(\gamma_t)}{u(\gamma_t)}-\frac{2 v'(\gamma_t)}{v(\gamma_t)}\right)\Bigg],
\end{aligned}$}
\end{equation}
with  $u'(\gamma_t)=\theta'_{\min }(\gamma_t)$ for $u(\gamma_t)\in\{\theta_{\min},\theta_v+\theta_{\min}\}$ and $u'(\gamma_t)=-\theta'_{\min }(\gamma_t)$ for $u(\gamma_t)=\theta_v-\theta_{\min}$, for which
\begin{align}
\theta'_{\min }(\gamma_t)=\begin{cases}
-\frac{\cos(\theta_c) \csc(\theta_v) \sec(\gamma_t) \tan(\gamma_t)}{1 + \left(\cot(\theta_v) - \cos(\theta_c) \csc(\theta_v) \sec(\gamma_t)\right)^2},
&\text{Case 2, }\\
\frac{\sin (\gamma_t) \sec (\theta_c) \csc(\theta_v)}{(\cot (\theta_v)-\cos (\gamma_t) \sec (\theta_c) \csc (\theta_v))^2+1}, &\text{Case 3, }\\
-\frac{\sin (\gamma_t) \sec (\theta_c) \csc(\theta_v)}{(\cot (\theta_v)-\cos (\gamma_t) \sec (\theta_c) \csc (\theta_v))^2+1}, &\text{Case 4, }\\
\end{cases}
\end{align}
whereas
\begin{align}
v'(\gamma_t) = 
\begin{cases} 
0, & \text{if } v(\gamma_t) = \theta_C, \\
1, & \text{if } v(\gamma_t) = \gamma_t. 
\end{cases}
\end{align}
In (\ref{eq:v_approx_der}), $_3F_2^{(0,0,0,0,0,1)}(\cdot,\cdot,\cdot;\cdot,\cdot;z)$ is the derivative of the hypergeometric function with respect to $z$ and can be calculated according to\cite[Eq.~07.31.20.0007.01]{wolframfunctions-2F3}.  }
\end{corollary}

The following Lemma is crucial for statistically characterizing the differential Doppler shift in Section~\ref{subs:Differential Doppler Shift}.

\begin{lemma}
\label{lemma:central_cdf}
\textit{Our extensive simulation results reveal that the CDF of the random variable $\Gammamin$, representing the minimum central angle of the uniformly distributed UEs within $S(C,\theta_c)$, can be approximated with a high level of accuracy by the CDF of the central angle $\Upupsilon_t$, given in (\ref{eq:central_cdf}), by substituting $\theta_v$ by the minimum central angle at $C$, i.e., $\mu_{_{\Gammamin}}$, as
\begin{align}
\label{eq:mean_cdf}
F_{_{\Gammamin}}(\gammamin) &= F_{\Upupsilon_t}(\gammamin,\theta_c\mid\mu_{_{\Gammamin}}),
\end{align}
and the PDF is thereof 
\begin{align}
\label{eq:theta_min_pdf}
f_{_{\Gammamin}}(\gammamin) &= f_{\Upupsilon_t}(\gammamin,\theta_c\mid\mu_{_{\Gammamin}}),
\end{align}
This result is validated in Fig.~\ref{fig:mingamma} by comparing the exact CDF of $\Gammamin$ obtained through Monte-Carlo simulations with the theoretical approximation given in (\ref{eq:mean_cdf}) for different system parameters, demonstrating an excellent match.}
\end{lemma}

\subsection{Statistical Analysis of Doppler Shift}
Leveraging the statistical measures derived in Section~\ref{subs:angle_distributions}, this subsection characterizes the Doppler shift induced by the LEO satellite on users within the cell.
\begin{theorem}
\label{theorem:cdf_doppler}  
\textit{The CDF of the Doppler shift magnitude within a spherical cap cell, centered at $C$ and separated by an angle $\theta_v$ from the serving satellite and bounded by an angle of $2\theta_c$, wherein the UEs are uniformly distributed, can be expressed as 
\begin{align}
\label{eq:CDF_DOPPLER_gen}
& F_{\delta}(s)=\E_{_{\Gammamin}}\Bigg[F_{\Upupsilon_t}\Bigg(\arccos{\bigg(\text{Cos}_{\Upsilon}(s) \bigg)},\theta_c\mid\theta_v\Bigg)\Bigg],
\end{align}
where
\begin{align}
\label{eq:cos_param}
&\text{Cos}_{\Upsilon}(s) 
&=\frac{s^2 k + \sqrt{s^4 k^2 - \rho^2 \big(s^2 (1 + k^2) - \rho^2 \cos^2(\Gammamin)\big)}}{\rho^2},
\end{align}
and $F_{\Upupsilon_t}(\cdot,\cdot\mid\cdot)$ is the CDF given in (\ref{eq:central_cdf}). The expectation in (\ref{eq:CDF_DOPPLER_gen}) is w.r.t. the minimum central angle of the UEs whose CDF and PDF are given in (\ref{eq:mean_cdf}) and (\ref{eq:theta_min_pdf}), respectively. 
}
\end{theorem}
\begin{IEEEproof}
The CDF of the Doppler shift magnitude $\delta_t$ defined in \eqref{eq:doppler_s} can be expressed as
\begin{align}
\label{eq:cdf_gen_derivation}
&F_{\delta}(s) = \operatorname{Pr}\big(\delta_t \leq s\big)\nonumber\\
&= \mathbb{E}_{\Gammamin}\left[\operatorname{Pr}\big(\delta_t \leq s \mid \Gammamin\big)\right]\nonumber\\
&\stackrel{}{=} \mathbb{E}_{\Gammamin}\left[\operatorname{Pr}\left(\rho \, \frac{\sqrt{\cos^2(\Gammamin) - \cos^2(\Upsilon_t)}}{\sqrt{1 + k^2 - 2k\cos(\Upsilon_t)}} \leq s \mid \Gammamin\right)\right]\nonumber\\
&\stackrel{(a)}{=} \mathbb{E}_{\Gammamin}\left[\operatorname{Pr}\left(\Upsilon_t \leq \arccos\big(\text{Cos}_{\Upsilon}(s)\big) \mid \Gammamin\right)\right],
\end{align}
where \(\text{Cos}_{\Upsilon}(s)\) is defined above in \eqref{eq:cos_param} and (a) follows from solving the inequality  for \(\Upsilon_t\).
In particular, beginning with the Doppler shift expression in \eqref{eq:doppler_s}, we define \(x = \cos(\Upsilon_t)\). Substituting this and squaring both sides to eliminate the square roots, we have
\begin{align}
\rho^2 \, \frac{\cos^2(\Gammamin) - x^2}{1 + k^2 - 2kx} \leq s^2.
\end{align}
Rearranging terms and expanding both sides, we obtain:
\begin{align}
\rho^2 x^2 - 2s^2 k x + s^2 (1 + k^2) - \rho^2 \cos^2(\Gammamin) \leq 0.
\end{align}
This is a quadratic inequality in \(x = \cos(\Upsilon_t)\). Solving it, and restricting \(x\) to the range corresponding to the physical constraints (i.e., \(0 \leq \cos(\Upsilon_t) \leq 1\)), we obtain
\begin{align}
&\cos(\Upsilon_t)\nonumber\\
&\leq \frac{s^2 k + \sqrt{s^4 k^2 - \rho^2 \big(s^2 (1 + k^2) - \rho^2 \cos^2(\Gammamin)\big)}}{\rho^2}.
\end{align}
By applying the inverse cosine function to isolate $\Upsilon_t$, as shown in~\eqref{eq:cdf_gen_derivation}, and using the definition of the CDF, $\Pr(X \leq x)$, we substitute the CDF of $\Upsilon_t$ to obtain~\eqref{eq:CDF_DOPPLER_gen}.
\end{IEEEproof}

Since the CDF in (\ref{eq:CDF_DOPPLER_gen}) can only be evaluated numerically, either using numerical integration or Monte Carlo methods through random sampling over $\Gammamin$ with the PDF in (\ref{eq:theta_min_pdf}), in the following theorem we propose a very accurate approximation for the CDF of the Doppler shift magnitude.

\begin{theorem}
\label{theorem:cdf_doppler}  
\textit{The CDF of the Doppler shift magnitude within a spherical cap cell compassed by an angle of $2\theta_c$ and centered at $C$ that is separated by an angle of $\theta_v$ from the serving satellite and have a minimum central angle $\mu_{_{\Gammamin}}$, can be approximated as 
\begin{equation}
\resizebox{1\hsize}{!}{$\begin{aligned} \label{eq:CDF_DOPPLER}
& F_{\delta}(s\mid\Gammamin= \mu_{_{\Gammamin}})\\
&=F_{\Upupsilon_t}\Bigg(\cos^{-1}{\bigg(\frac{s^2 k + \sqrt{s^4 k^2 - \rho^2 \big(s^2 (1 + k^2) - \rho^2 \cos^2(\mu_{_{\Gammamin}})\big)}}{\rho^2}\bigg)},\theta_c\mid\theta_v\Bigg).
\end{aligned}$}
\end{equation}
}
\end{theorem}

\begin{figure}
    \centering
     \includegraphics[trim=.1cm .1cm .1cm .1cm, clip=true, width=0.48\textwidth ]{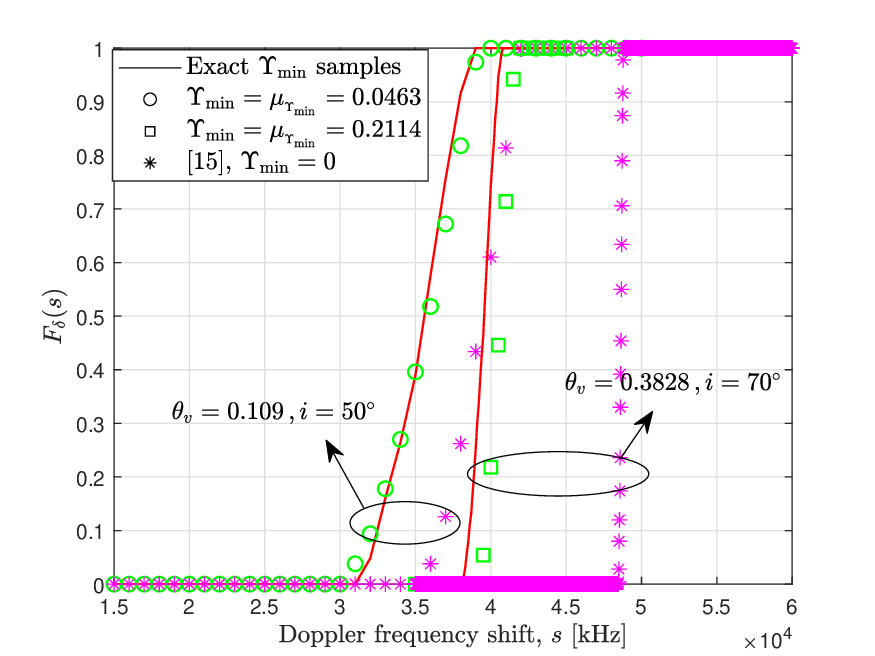}
    \caption{ Validation of the constant minimum central angle approximation introduced in Theorem \ref{theorem:cdf_doppler}: Comparison between the simulated CDF of the Doppler shift using (i) exact $\Gammamin$ samples, (ii) the proposed constant $\mu_{_{\Gammamin}}$, and (iii) the approach in \cite{main} ignoring $\Gammamin$.}
    \label{fig:validate_constant_angle}
\end{figure}

\begin{IEEEproof}
Using the observation that rather than averaging over the random variable $\Gammamin$, an accurate approximation is achieved by directly substituting the constant minimum central angle at the cell's center $,\mu_{_{\Gammamin}}$, in place of $\Gammamin$, in such a way that
\begin{equation}
\resizebox{1\hsize}{!}{$\begin{aligned}   &  F_{\delta}(s\mid\Gammamin= \mu_{_{\Gammamin}}){=}\operatorname{Pr}\Bigg(\Upupsilon_t\leq\arccos\bigg(\frac{s^2 k}{\rho^2}\nonumber\\
& + \frac{\sqrt{s^4 k^2 - \rho^2 \big(s^2 (1 + k^2) - \rho^2 \cos^2(\mu_{_{\Gammamin}})\big)}}{\rho^2}\bigg)  \mathrel{\Big|}  \Gammamin= \mu_{_{\Gammamin}}\Bigg).
\end{aligned}$}
\end{equation}
This approach simplifies the analytical results, avoiding any further complexity. The validation of the employed approach is demonstrated through Monte-Carlo simulations, as depicted in Fig.~\ref{fig:validate_constant_angle}, wherein the simulated CDF of the Doppler shift magnitude, calculated using the actual $\Gammamin$ samples, is compared with the simulated CDF calculated using $\mu_{_{\Gammamin}}$. Additionally, the simulated CDF resulting from completely disregarding $\Gammamin$, as done in \cite{main}, is included for comparison. This comparison serves to underscore the significant importance of incorporating $\Gammamin$ in the analysis, highlighting the vulnerability of the analysis done in \cite{main}. Our approach significantly outperforms that in \cite{main} by achieving near-perfect alignment between the exact CDF and its approximation, which is based on a constant minimum central angle for all UEs within the cell at a specific time instant. 
\end{IEEEproof}

\begin{corollary}
\label{corol:pdf_doppler}  
\textit{The PDF of the Doppler shift magnitude within a spherical cap cell that is enclosed by an angle $2\theta_c$, wherein the UEs are uniformally distributed, can be expressed as (\ref{eq:PDF_DOPPLER}) at the beginning of the previous page, 
for which $f_{\Upupsilon_t}(\cdot,\cdot\mid\cdot)$ is the PDF given in (\ref{eq:central_angle_pdf}) of the corresponding central angle $\Upupsilon_t$. The PDF in (\ref{eq:PDF_DOPPLER}) follows directly from differentiating (\ref{eq:CDF_DOPPLER}) with respect to the parameter $s$ 
using the chain rule.\stepcounter{equation}}
\end{corollary}

\begin{corollary}
\label{corol:CDF_DOPPLER_approx}
\textit{The CDF of the Doppler shift magnitude within a spherical cap cell that is enclosed by an angle $2\theta_c$, wherein the UEs are uniformly distributed, can be upper bounded by
\begin{align}
\label{eq:CDF_upper_bound}
& \widetilde{F}_{\delta}(s){=}F_{\Upupsilon_t}\left({\frac{\big(\frac{h}{\re+h}\big)\,s}{\sqrt{\rho^2-\big(\frac{\re}{\re+h}\big)\,s^2}}},\theta_c\mid\theta_v\right),
\end{align}
where $F_{\Upupsilon_t}(\cdot,\cdot\mid\cdot)$ is given in (\ref{eq:central_cdf})}.
\end{corollary}
\begin{IEEEproof}
As per \cite[ Lemma 2]{main}, the Doppler shift at a UE with a minimum central angle $0<\Gammamin\leq\phimax$, is bounded by the Doppler shift at a UE with a minimum central angle $\Gammamin=0$. Accordingly, the Doppler shift expression in~\eqref{eq:doppler_s} is upper-bounded by the simplified form in~\eqref{eq:doppler_s_approx_theta_min_0}. Therefore, substituting (\ref{eq:doppler_s_approx_theta_min_0}) into the CDF expression $\operatorname{Pr}\big({\widetilde{\delta}_t} \leq s\big)$ and applying algebraic simplifications, yields the upper bound CDF in~\eqref{eq:CDF_upper_bound}. 
\end{IEEEproof}

\subsection{Statistical Analysis of the Differential Doppler Shift}
\label{subs:Differential Doppler Shift}
In wireless communication systems, Doppler compensation plays a crucial role in mitigating the effects of Doppler resulting from the movement of satellites with respect to user terminals. A reference point, usually the center of the beam, is chosen within each spherical cap cell to offset the common component of the Doppler shift experienced by all its users. Once the Gateway calculates the common time-varying Doppler shift at the reference point, this component can be pre-compensated in the downlink or post-compensated in the uplink, effectively nullifying the Doppler shift at that reference point. Consequently, the residual Doppler shift experienced by UEs in the cell depends on their relative position to the reference point.

\begin{theorem}
\label{theor:DD_GEN}
For a reference point $C$ in $S(C,\theta_c)$, with central angle $\theta_v$ to the serving satellite and a minimum central angle $\mu_{_{\Gammamin}}$, the CDF and PDF of the differential Doppler shift within the spherical cap cell can be calculated respectively as 
\begin{align}
\label{eq:CDF_DD}
F_{_{\Delta \delta}}(\zeta)&{=}F_{\delta}\Bigg(\zeta+ \rho \, \frac{\sqrt{\cos^2(\mu_{_{\Gammamin}}) - \cos^2(\theta_v)}}{\sqrt{1 + k^2 - 2k\cos(\theta_v)}}\Bigg),
\end{align}
and
\begin{align}
\label{eq:PDF_DD}
f_{_{\Delta \delta}}(\zeta)&{=}f_{\delta}\Bigg(\zeta+ \rho \, \frac{\sqrt{\cos^2(\mu_{_{\Gammamin}}) - \cos^2(\theta_v)}}{\sqrt{1 + k^2 - 2k\cos(\theta_v)}}\Bigg),
\end{align}
for which $F_{\delta}(\cdot)$ is defined in (\ref{eq:CDF_DOPPLER_gen}) or (\ref{eq:CDF_DOPPLER}), whereas $f_{\delta}(\cdot)$ is defined in (\ref{eq:PDF_DOPPLER}). 
\end{theorem}
\begin{IEEEproof}
The Doppler shift at any UE in $S(C,\theta_c)$ can be written as $\delta_t=D_t+\Delta \delta_t$, for which $D_t$ is the common part of Doppler shift experienced by all the UEs in the cell and is calculated using (\ref{eq:doppler_s}) at the reference point with substituting $\Upupsilon_t=\theta_v$ and $\Gammamin=\mu_{_{\Gammamin}}$. Therefore, $\operatorname{Pr}\big(\Delta \delta_t \leq \zeta\big)= \operatorname{Pr}\big(\delta_t-D_t\leq \zeta\big)=F_{\delta}(\zeta+D_t)$ as in (\ref{eq:CDF_DD}). The PDF in (\ref{eq:PDF_DD}) is obtained by differentiating the corresponding CDF in (\ref{eq:CDF_DD}) with respect to $\zeta$.
\end{IEEEproof}

\begin{figure}
\begin{center}
\subfigure[]{\includegraphics[trim=1.80cm 5cm 4.5cm .1cm, clip=true, width=.49\textwidth]{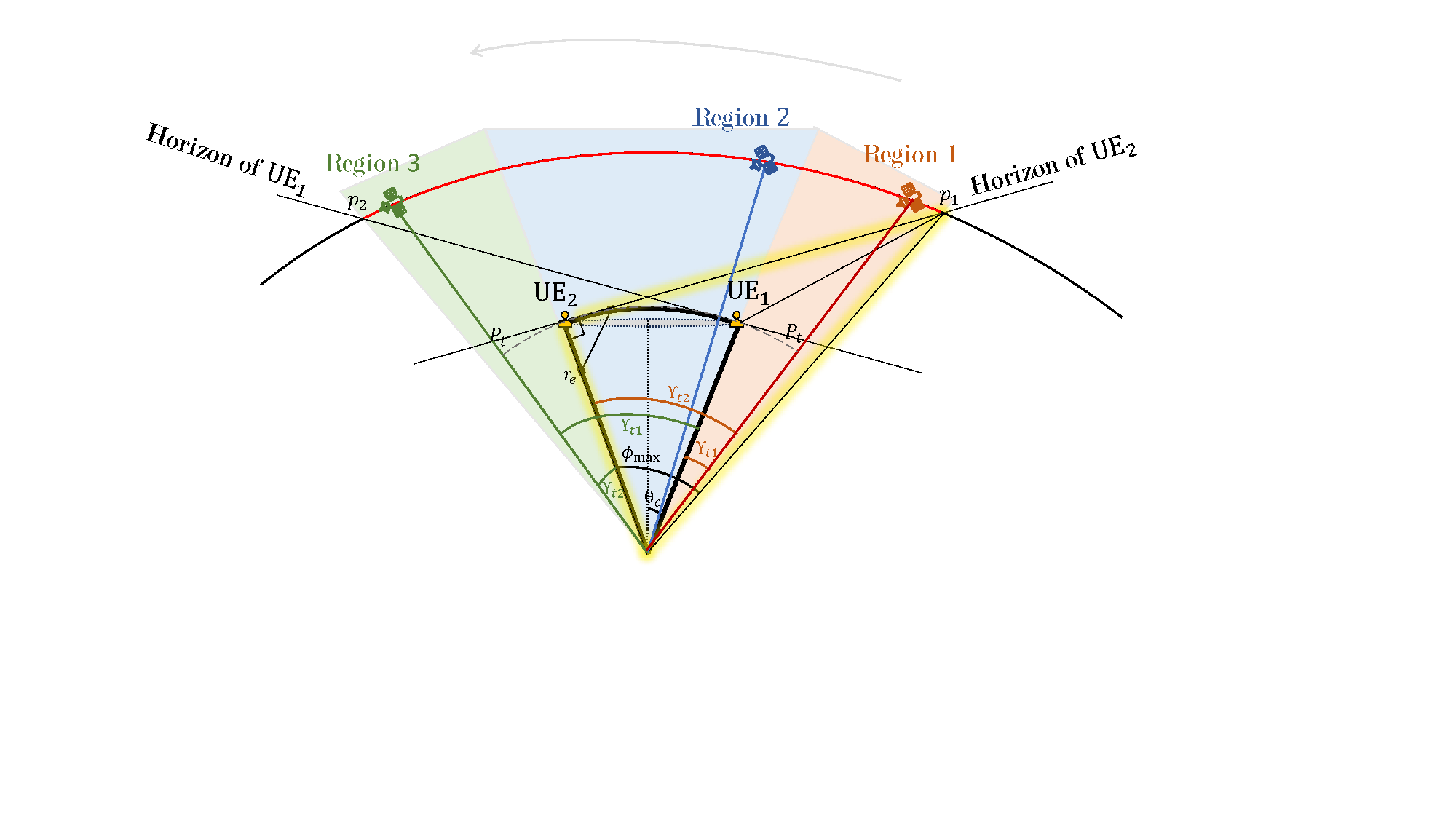}}
\subfigure[]{\includegraphics[trim=1.80cm 5cm 4.5cm .1cm, clip=true, width=.49\textwidth]{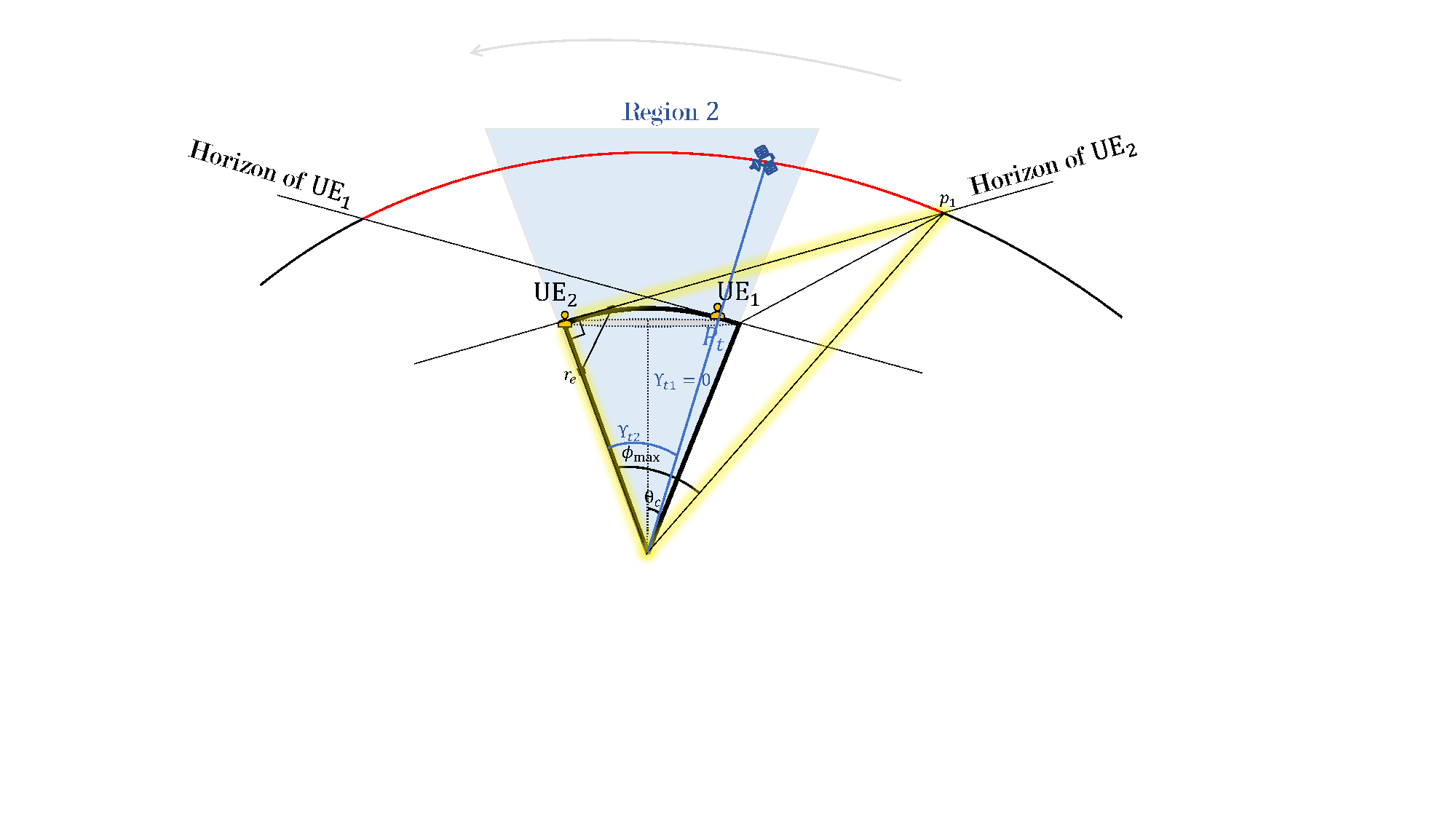}}
\caption{Geometric illustration used to analyze and derive the maximum differential Doppler shift within a spherical cap cell containing uniformly distributed UEs. (a) Worst-case maximum Doppler shift difference in Scenario 2 when \( P_t \) lies outside the coverage cell \( S(C, \theta_c) \), occurring between two UEs separated by angle \( 2\theta_c \) along the satellite ground track. (b) Maximum Doppler shift difference in Scenario 1 when \( P_t \) lies inside the coverage cell \( S(C, \theta_c) \), occurring between the UE aligned with \( P_t \) and a UE at the cell's edge. }
\label{fig:maximum_DD}
\end{center}
\end{figure}

\subsection{Maximum Differential Doppler Characterization}
\label{Maximum Differential Doppler Characterization}
In this subsection, we identify the maximum differential Doppler shift in a spherical cap cell that contains uniformly distributed UEs at random. Since the Doppler shift magnitude is written in closed form in terms of the central angle in (\ref{eq:doppler_s}), it is straightforward to find the exact maximum differential Doppler shift, opposing to the analysis in~\cite{dop_iot}, which is based upon simplified assumptions including flat-Earth approximation with 2D plane instead of a spherical surface. 

Before defining the maximum differential Doppler within a cell, it is essential to ensure that all UEs in the cell share the same satellite visibility duration. This implies that, at any given time instant during the satellite's visibility window, all UEs can be served by the same satellite. To achieve this, the maximum common satellite visibility angular region, referenced to the center of the cell and denoted by ${\phi^\text{com}_{\text{max}}}$, is calculated. Consequently, the visible satellite for all users is assumed to remain within this angular region.
This condition is directly derived from the geometry illustrated in Fig.~\ref{fig:maximum_DD}(a), where the Earth's radius, extending from a specific UE (e.g., UE$_2$) to the Earth's center, the distance from the UE to the horizon intersection with the satellite's orbit (\(p_1\) for UE$_2$), and the distance from the intersection point ($p_1$) to the Earth's center form a right triangle. From this geometry, the maximum central angle for any UE, \(\phi_{\text{max}}\), can be computed as $\phimax = \arccos\left(\frac{\re}{\re + h}\right)$. The common visible horizon for all UEs is depicted by the red curve in Fig.~\ref{fig:maximum_DD}, extending from $p_1$ to $p_2$. These points correspond to the furthest positions along the satellite’s orbit where it remains simultaneously visible to all UEs within the coverage cell. Specifically, $p_1$ is the farthest point visible to UE$_2$, and $p_2$ to UE$_1$. While UE$_1$ and UE$_2$ define the cell's edges, $p_1$ and $p_2$ outline the boundary of the cell’s visible horizon. Beyond this boundary, the satellite is no longer visible to all UEs, and thus cannot serve the entire cell. The critical angle, \(\phi^\text{com}_{\text{max}}\), is therefore defined by

\begin{align}
    \phi^\text{com}_{\text{max}} = \phimax - \theta_c.
\end{align}

\begin{theorem}
\label{theor:MDDS}
\textit{The maximum differential Doppler shift at an instant of time within a spherical cap cell containing uniformly distributed UEs at random is given by
\begin{align}
\label{eq:MDDS}
&\Delta\delta_{\text{max},t}(\theta_v)=\abs{{\delta}_t (\Upupsilon_{t_2})-{\delta}_t (\Upupsilon_{t_1})}\nonumber\\
&=\rho\,\Bigg|\Bigg( \,\frac{\sin\big(\Upupsilon_{t_2}(\theta_v)\big)}{\sqrt{1+k^2-2k\cos\big(\Upupsilon_{t_2}(\theta_v)\big)}}\nonumber\\
&-\,\frac{\sin\big(\Upupsilon_{t_1}(\theta_v)\big)}{\sqrt{1+k^2-2k\cos\big(\Upupsilon_{t_1}(\theta_v)\big)}}\Bigg)\Bigg|,
\end{align}
where 
\begin{equation}
\resizebox{.98\hsize}{!}{$\begin{aligned}    \Upupsilon_{t_2}(\theta_v)\hspace{-.1 cm}=\hspace{-.1 cm}\begin{cases}
       \theta_c+\theta_v, \text{region 1: $P_t$ outside \& approaches cell}\\
     \theta_c+\theta_v, \text{region 2: $P_t$ inside cell} \\
     \theta_v-\theta_c, \text{region 3: $P_t$ outside \& recedes cell,}\nonumber
    \end{cases}
\end{aligned}$}
\end{equation}
and 
\begin{equation}
\resizebox{.98\hsize}{!}{$\begin{aligned}    \Upupsilon_{t_1}(\theta_v)\hspace{-.1 cm}=\hspace{-.1 cm}\begin{cases}
       \theta_v-\theta_c,& \text{region 1: $P_t$ outside \& approaches cell}\\
     0, &\text{region 2: $P_t$ inside cell} \\
     \theta_v+\theta_c,& \text{region 3: $P_t$ outside \& recedes cell.}\nonumber
    \end{cases}
\end{aligned}$}
\end{equation}}
\end{theorem}
\begin{IEEEproof}
In Scenario 2, where \( P_t \) lies outside the coverage cell \( S(C, \theta_c) \), the {maximum Doppler shift difference} within the cell occurs between two UEs located at the maximum angular separation of \( 2\theta_c \) along the satellite’s ground track. This \emph{worst-case scenario} occurs between UE$_1$ and UE$_2$ in Fig.~\ref{fig:maximum_DD}(a), where \( \Gammamin = 0 \).
Therefore, the Doppler shift at UE$_1$ and UE$_2$ that has central angles $\Upupsilon_{t_1}$ and $\Upupsilon_{t_2}$, respectively, to the serving satellite, can be expressed as ${\delta}_t (\Upupsilon_{t_1})$ and ${\delta}_t (\Upupsilon_{t_2})$, where ${\delta}_t (\cdot)$ is defined in (\ref{eq:doppler_s_simplified}). Since the maximum differential Doppler have the form $\Delta\delta_{\text{max},t}=\abs{{\delta}_t (\Upupsilon_{t_2})-{\delta}_t (\Upupsilon_{t_1})}$, we need to write the central angles $\Upupsilon_{t_2}$ and $\Upupsilon_{t_1}$  in terms of $\theta_c$ and $\theta_v$ which are already defined for the Gateway. From the geometry illustrated in Fig.~\ref{fig:maximum_DD}(a), we can conclude that when $P_t$ is outside $S(C,\theta_c)$, $\Upupsilon_{t_2}=2\,\theta_c+\Upupsilon_{t_1}$, and $\Upupsilon_{t_1}=\theta_v-\theta_c$ when the satellite approaches the cell, whereas $\Upupsilon_{t_2}=\Upupsilon_{t_1}-2\,\theta_c$, and $\Upupsilon_{t_1}=\theta_c+\theta_v$ when the satellite moves away from the cell. On the other hand, for scenario 1, where $P_t$ lies inside $S(C,\theta_c)$, the maximum difference in Doppler shift in the cell can occur between two UEs from which one of them aligns with the point $P_t$ and the other is on the edge of $S(C,\theta_c)$ as illustrated in Fig.~\ref{fig:maximum_DD}(b) with $\Gammamin=0$. Therefore, $\Upupsilon_{t_1}=0$ and $\Upupsilon_{t_2}=\theta_c+\theta_v$.\end{IEEEproof}


\begin{figure}[t]
    \centering
     \includegraphics[trim=.1cm .1cm .1cm .1cm, clip=true, width=0.48\textwidth ]{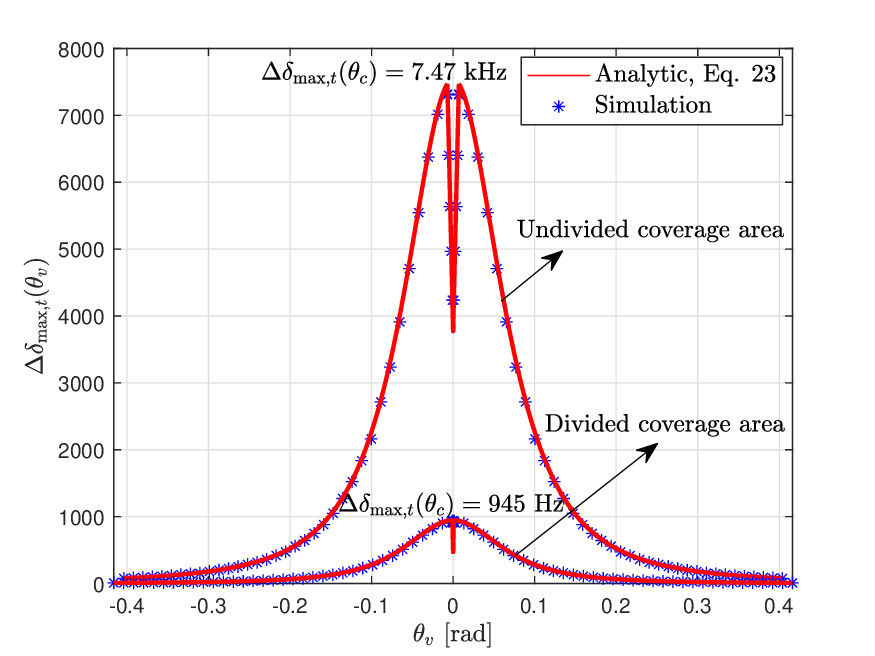}
    \caption{ Maximum differential Doppler shift $\Delta\delta_{\text{max},t}(\theta_v)$ versus $\theta_v$ for a satellite at an altitude of 600 km with carrier frequency $f_o=2$ GHz. The upper curve corresponds to the original cell with $90$ km diameter, while the lower curve illustrates the reduced maximum differential Doppler shift after subdividing the cell into smaller regions with $11.25$ km diameter to meet the allowable threshold of 950 Hz.}
    \label{fig:numeric_max_DD}
\end{figure}

\subsection{Reduction of Differential Doppler Shift}
\label{subs:Reduction of Differential Doppler Shift}
To mitigate the effects of differential Doppler among UEs, one effective approach is to group them into smaller clusters within the cell based on their locations. Frequency compensation techniques are then applied at a reference point within each cluster, resulting in decreased differential Doppler among UEs within the same cell. In order to efficiently group the random UEs inside the cell, we set an upper threshold on the allowable differential Doppler shift for each cluster in the cell using the maximum differential Doppler shift defined in (\ref{eq:MDDS}), after which $\theta_c$ that achieves this threshold can be directly calculated. A simple showcase is provided in the following example.
\begin{example}
For a satellite at altitude $h=600$ km that forms a cell on the Earth's surface with a spherical radius $\theta_c=0.0071$ rad (corresponds to a diameter of length $90$ km), the maximum differential Doppler shift in the spherical cap cell that contains uniformly distributed UEs at random is plotted in Fig.~\ref{fig:numeric_max_DD} for $f_o=2$ GHz as a function of $\theta_v$ using (\ref{eq:MDDS}).
The figure shows that the peak value of the maximum differential Doppler shift for a cell of UEs occurs when the satellite's projection is at the cell's edge ($\theta_v=\theta_c$). Notably, at $\theta_v = 0$, the maximum differential Doppler shift for the given system setup is $3760$ Hz, as shown in Fig.~\ref{fig:numeric_max_DD}. This value is consistent with the maximum Doppler shift reported in \cite[Table 6.1.1.1-8]{specs_doppler} for the same configuration, where a shift of $1.88$ ppm (approximately $3760$ Hz) is observed, assuming a pre/post-compensation mechanism is applied at a reference point.
According to \cite{dop_iot}, the 3GPP specification supports differential Doppler shifts of up to $950$ Hz among mobile UEs in terrestrial networks, where users may travel at speeds up to $500$ km/h under comparable system conditions. For LEO satellite communications, UEs are generally expected to accommodate higher maximum differential Doppler shifts, typically ranging from several hundred to a few kilohertz, because of the much higher relative velocities involved.
To ensure that the differential Doppler shift within each coverage region remains below a specified threshold, such as the $950$ Hz limit, the coverage cell is subdivided into smaller clusters. In the example considered, the total coverage area, given by $2\pi \re^2(1 - \cos(0.0071))$, is divided into smaller circular beams. Each beam has an angular radius of $\theta_c = 8.83 \times 10^{-4}$, determined numerically using~(\ref{eq:MDDS}) to ensure that the maximum differential Doppler shift within a beam does not exceed the threshold. This results in $\frac{(1 - \cos(0.0071))}{(1 - \cos(8.83 \times 10^{-4}))}=64$ clusters, obtained by dividing the total cell's area by the area of each smaller beam.
Each beam covers a ground-projected diameter of approximately $11.25$ km, forming realistic circular footprints that are practically achievable using frequency reuse techniques. This contrasts with the partitioning approach in \cite{dop_iot}, which divides the area into vertical segments that do not align with the physical characteristics of actual satellite antenna patterns. As shown by the lower curve in Fig.~\ref{fig:numeric_max_DD}, this subdivision effectively limits the differential Doppler shift within each beam, with a maximum value of $\Delta\delta_{\text{max},t}(\theta_v) = 945$ Hz occurring at $\theta_v = \theta_c$.
\end{example}

\section{Numerical Results and Discussion}
\label{sec:numerical_results}

This section validates the analytical expressions derived in Section~\ref{sec:doppler characterization} through extensive system-level simulations. It also provides a comprehensive elaboration of the impact of various system parameters on the CDF of the Doppler shift magnitude at a random UE within the cell or the smaller clusters.
We consider herein a network setup with the parameters listed in Table~\ref{table:pars}.
To generate the numerical results, we simulated the Walker-star constellation, which is composed of several orbital planes that are uniformly rotated around the Earth's $z$-axis with equal spacing, and each orbit contains the same number of satellites that are equally spaced. It is worth mentioning that while Table~\ref{table:pars} presents baseline parameter values based on the 3GPP technical report \cite{specs_doppler}, alternative values can be effectively tested to evaluate different scenarios and system behaviors.

\begin{table}[h!]
    \centering
    \resizebox{\linewidth}{!}{
        \begin{tabular}{lll}
            \hline \hline
            Symbol & Parameter & Value \\
            \hline
            $\re$ & Earth's radius & $6,371 \mathrm{~km}$ \\
            $h$ & Satellite altitude & $600,\,1200,\,2000 \mathrm{~km}$ \\
            $f_{\mathrm{o}}$ & Carrier frequency & $2$ GHz\\
            $c$ & Speed of light & $3 \times 10^{8}$ m/sec\\
                       $\mu$ & Standard gravitational parameter of Earth & $3.986\times 10^{14}$ m$^{3}$/sec$^2$\\
            $\omega_E$ & Earth's angular velocity in ECI & $7.27 \times 10^{-5}$ rad/sec \\
            $i$ & Constellation's inclination angle & $53^{\circ}$ \\
            $\theta_c$ & Angular radius of the cell & $0.0078$ rad ($R_{\text{cell}}=50$ km) \\
            \hline
        \end{tabular}
    }
    \caption{Parameter values}
    \label{table:pars}
\end{table}
\begin{figure}[]
    \centering
     \includegraphics[trim=.1cm .1cm .1cm .1cm, clip=true, width=0.48\textwidth ]{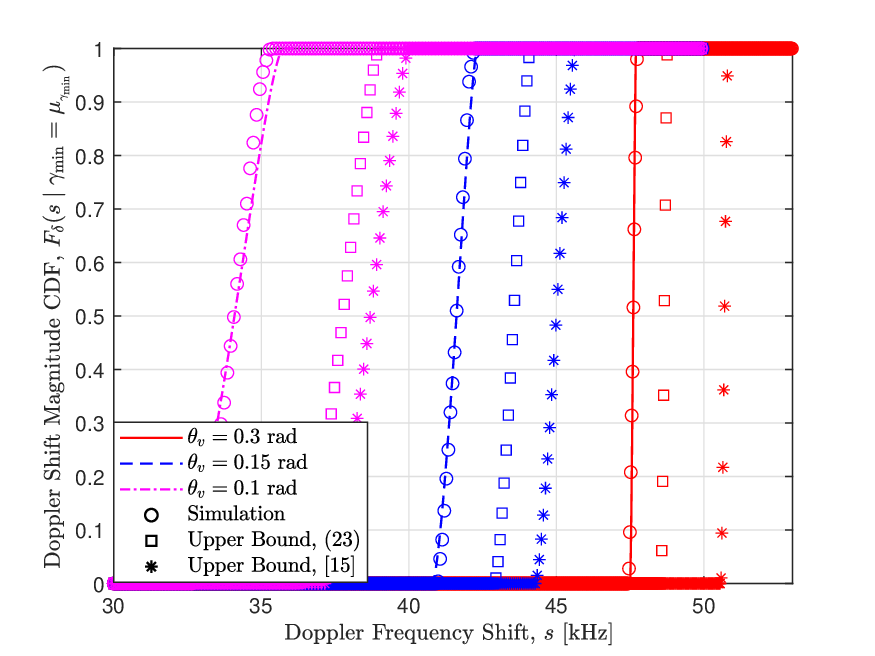}
    \caption{ Comparison of the exact CDF of the Doppler shift magnitude (Monte Carlo simulation) and the proposed CDF derived in (\ref{eq:CDF_DOPPLER}) for varying satellite positions. The figure also compares the proposed tighter upper bound (\ref{eq:CDF_upper_bound}) with the bound from \cite{main}. }
    \label{fig:cdf_thetav}
        \vspace{-.5cm}
\end{figure}

\begin{figure}[]
    \centering
     \includegraphics[trim=.1cm .1cm .1cm .1cm, clip=true, width=0.48\textwidth ]{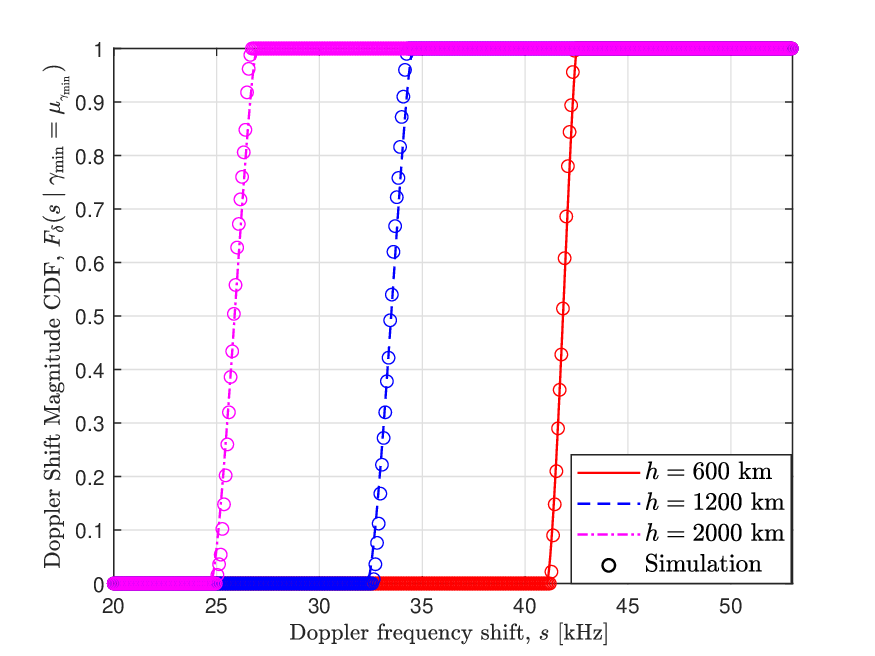}
    \caption{Impact of satellite altitude on the CDF of the Doppler shift magnitude.}
    \label{fig:cdf_altitude}
    \vspace{-.5cm}
\end{figure}

We begin by demonstrating the high accuracy of the proposed mathematical approach by comparing the exact CDF of the Doppler shift magnitude, calculated via Monte Carlo simulation, with the theoretical CDF derived in (\ref{eq:CDF_DOPPLER}), as illustrated in Fig. \ref{fig:cdf_thetav}. This comparison is carried out for a cluster with an angular radius of $\theta_c =0.0078$ rad, which corresponds to a cell's radius of $R_{\text{cell}}=50$ km, and a minimum central angle $\mu_{_{\Gammamin}} =0.042$ rad. The evaluation is conducted at multiple time instants, each corresponding to a different value of $\theta_v$ within the satellite visibility window, i.e, different satellite
locations relative to the cell's center. The results, depicted in Fig. \ref{fig:cdf_thetav}, demonstrate a remarkable agreement between the theoretical and simulated CDFs, providing strong validation of the derived analytical CDF. A rightward shift in the CDF is noted, which implies that higher Doppler shifts become more probable as the satellite moves away from the center. This is consistent with the general Doppler effect behavior, where the maximum Doppler shift is observed near the horizon, while it is lower when the satellite is directly overhead. The comparison is also benchmarked against the upper bound derived in \cite{main}, which reveals that this bound is very loose and can deviate significantly from the exact CDF. To address this limitation, a tighter upper bound is developed in this work, as given in (\ref{eq:CDF_upper_bound}), and is also plotted in the figure, showing a substantially improved fit compared to the bound proposed in \cite{main}, particularly at the instants when the satellite is momentarily away from the cell's center.

The effect of satellite altitude on the magnitude of the Doppler shift is illustrated in Fig. \ref{fig:cdf_altitude}. As observed, higher satellite altitudes lead to an improvement in the Doppler CDF, corresponding to a reduction in the induced Doppler shift. This observation can be attributed to the decrease in relative velocity between the satellite and the ground user as the altitude increases, resulting in smaller Doppler variations. Furthermore, the simulation results show an excellent agreement with the analytical curves, confirming the accuracy of the proposed analytical model.

\begin{figure}[h!]
    \centering
     \includegraphics[trim=.1cm .1cm .1cm .1cm, clip=true, width=0.48\textwidth ]{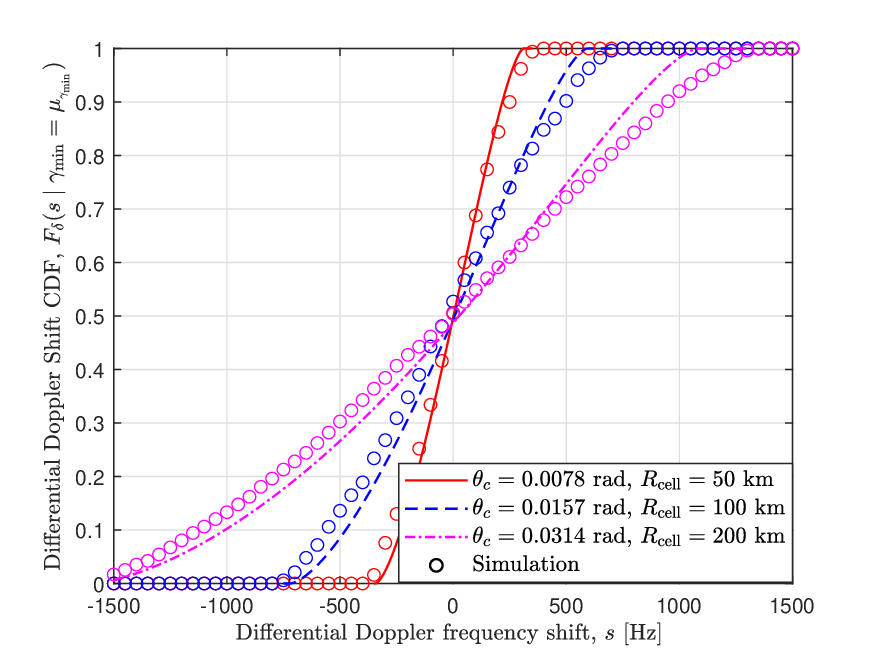}
\caption{CDF of the differential Doppler frequency shift for different spot beam coverage angles $\theta_c$ (corresponding to cell radii $R_{\text{cell}} = 50$, $100$, and $200$ km).}
    \label{fig:cdf_cell_size}
        \vspace{-.5cm}
\end{figure}

Fig~\ref{fig:cdf_cell_size} illustrates the CDF of the differential Doppler shift within a spherical cap cell, derived in~\eqref{eq:CDF_DD}, for various beam coverage angles $\theta_c$, corresponding to different cell radii. The results show strong agreement between the analytical expression and simulation across all cases, including for larger values of $\theta_c$, which represent broader spot beam coverage on the Earth's surface. It is also observed that increasing $\theta_c$ leads to a wider spread of differential Doppler shifts, highlighting the trade-off between coverage area and Doppler variation.

\begin{figure}[h!]
    \centering
     \includegraphics[trim=.1cm .1cm .1cm .1cm, clip=true, width=0.48\textwidth ]{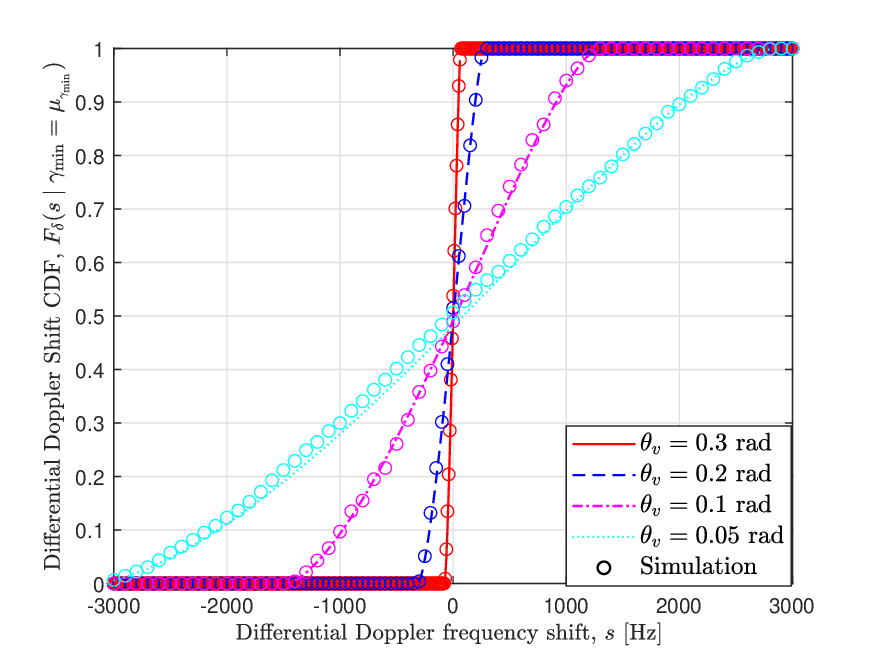}
    \caption{Impact of satellite position on the CDF of the differential Doppler frequency shift.}
    \label{fig:cdf_diff_theta_v}
        \vspace{-.5cm}
\end{figure}

While the impact of satellite location on the CDF of the Doppler shift magnitude is illustrated in Fig.~\ref{fig:cdf_thetav}, indicating that higher Doppler shifts are more likely when the satellite is positioned farther from the cell's center, Fig.~\ref{fig:cdf_diff_theta_v} explores its effect on the CDF of the differential Doppler frequency shift. As the central angle $\theta_v$ increases, corresponding to the satellite being farther from the edge of the cell, the spread of differential Doppler shifts narrows. Conversely, when the satellite is near the edge of the cell (i.e., smaller $\theta_v$), the spread becomes wider. This inverse relationship aligns with the trend shown in Fig.~\ref{fig:numeric_max_DD}, where the maximum differential Doppler shift peaks when the satellite is directly above the cell's edge. These findings emphasize the critical role of the position of the satellite in the determination of Doppler characteristics. Specifically, when the satellite is farther from the center, the Doppler shift experienced by the user is higher, while the variation in Doppler shifts across different users in the cell becomes smaller.





\section{Conclusion}
This paper presented a rigorous analytical framework for characterizing the Doppler shift and the differential Doppler shift in LEO satellite systems. By modeling the satellite beam footprint as a spherical cap and assuming uniformly distributed UEs, we derived closed-form expressions for the PDF and CDF of the Doppler shift magnitude at a random UE, along with an accurate upper bound for the Doppler shift distribution. Additionally, we developed analytical expressions for the PDF and CDF of the differential Doppler shift within the cell, providing key insights into the intra-cell Doppler variability resulting from user distribution and satellite movement.
We also showed that appropriately clustering users based on differential Doppler thresholds can effectively mitigate synchronization challenges in LEO networks.
The accuracy of the proposed analytical models was validated through extensive Monte Carlo simulations over realistic LEO constellations. The results demonstrated a strong agreement between the theoretical and simulated Doppler CDFs under various satellite altitudes, beam sizes, and user locations. Notably, our analysis revealed that when the satellite is further from the center of the cell, the Doppler shift increases due to higher relative velocity, while the variation in Doppler shifts across users becomes narrower. Furthermore, increasing satellite altitude reduces the Doppler shift magnitude, and larger beam coverage angles result in a broader spread of differential Doppler shifts. This work lays the groundwork for optimizing synchronization and resource allocation in LEO satellite networks by rigorously characterizing Doppler shift and differential Doppler shift. Our analysis enables precise prediction of Doppler-induced effects, facilitating more effective user clustering, enhanced Doppler compensation, and optimized satellite footprint design. These advancements can collectively minimize Doppler-induced interference, improving the reliability and efficiency of next-generation LEO satellite communication systems.
\bibliographystyle{IEEEtran}
\bibliography{refs}

\end{document}